\documentclass[twocolumn, prb, superscriptaddress]{revtex4-1}

\usepackage{graphicx}
\usepackage{amsmath}
\usepackage{hyperref}

\newcommand{\mof}{MOF-74}
\newcommand{\reaction}{H$_2$O~$\rightarrow$~OH+H}
\newcommand{\reactionD}{D$_2$O~$\rightarrow$~OD+D}

\graphicspath{{./figures/}}
\newcommand{\atomA}{{\bfseries\sffamily A}}
\newcommand{\atomB}{{\bfseries\sffamily B}}

\newcommand{\UTD}{Department of Materials Science and Engineering,
University of Texas at Dallas, Richardson, Texas 75080, USA}
\newcommand{\WFU}{Department of Physics, Wake Forest University,
Winston-Salem, North Carolina 27109, USA} 
\newcommand{\RU}{Department of
Chemistry and Chemical Biology, Rutgers University, Piscataway, New
Jersey 08854, USA}

\begin{document}

\title{Cluster Assisted Water Dissociation
Mechanism\\ in \mof\ and Controlling it Using Helium}

\author{S. Zuluaga}                               \affiliation{\WFU}
\author{E. M. A. Fuentes-Fernandez}               \affiliation{\UTD}
\author{K. Tan}                                   \affiliation{\UTD}
\author{J. Li}                                    \affiliation{\RU}
\author{I. J. Chabal}                             \affiliation{\UTD}
\author{T. Thonhauser} \email{thonhauser@wfu.edu} \affiliation{\WFU}

\date{\today}

\begin{abstract}
We show that the water dissociation reaction {\reaction} in the confined
environment of {\mof} channels can be precisely controlled by the
addition of the noble gas He. Elucidating the entire reaction process
with \emph{ab initio} methods and infrared (IR) spectroscopy, we prove
that the interaction between water molecules is critical to the formation
of water clusters, which reduce the dissociation barrier by up to 37\%
and thus influence the reaction significantly. Our time-resolved IR
measurements confirm that the formation of these clusters can be
suppressed by introducing He gas, providing unprecedented control over
water dissociation rates. Since the water dissociation reaction is the
cause of the structural instability of {\mof} in the presence of water,
our finding of the reaction mechanism lays the groundwork for designing
water stable versions of \mof\ as well as understanding water related
phenomena in MOFs in general.
\end{abstract}

\maketitle

\section{Introduction}

Metal organic framework (MOF) materials in general---and {\mof}
[\ensuremath{\mathcal{M}}$_2$(dobdc), \ensuremath{\mathcal{M}} =
Mg$^{2+}$, Zn$^{2+}$, Ni$^{2+}$, Co$^{2+}$, and
dobdc=2,5-dihydroxybenzenedicarboxylic acid] with its high density of
coordinatively unsaturated metal ions in particular---have shown
promising properties for technologically critical applications such as
gas storage and sequestration,\cite{Murray_2009:hydrogen_storage,
Li_2011:carbon_dioxide, Qiu_2009:molecular_engineering,
Nijem_2012:tuning_gate, Lee_2015:small-molecule_adsorption}
catalysis,\cite{Lee_2009:metal_organic, Luz_2010:bridging_homogeneous}
polymerization,\cite{Uemura_2009:polymerization_reactions,
Vitorino_2009:lanthanide_metal}
luminescence,\cite{Allendorf_2009:luminescent_metal,
White_2009:near-infrared_luminescent} non-linear
optics,\cite{Bordiga_2004:electronic_vibrational} magnetic
networks,\cite{Kurmoo_2009:magnetic_metal-organic} targeted drug
delivery,\cite{Horcajada_2010:porous_metal-organic-framework}
multiferroics,\cite{Stroppa_2011:electric_control,
Stroppa_2013:hybrid_improper, Di-Sante_2013:tuning_ferroelectric} and
sensing.\cite{Serre_2007:role_solvent-host,
Allendorf_2008:stress-induced_chemical, Tan_2011:mechanical_properties,
Kreno_2012:metal-organic_framework, Canepa_2015:structural_elastic} This
versatility is the result of MOF's simple nano-porous building-block
nature and affinity towards small molecule adsorption.  Unfortunately,
despite its great potential, the practical use of {\mof} in such
applications is rather limited due to its low stability and reduction of
gas uptake capacity in the presence of water
vapor.\cite{Schoenecker_2012:effect_water, DeCoste_2013:effect_water,
Kizzie_2011:effect_humidity, Remy_2013:selective_dynamic,
Liu_2011:stability_effects}

\begin{figure}
\includegraphics[width=\columnwidth]{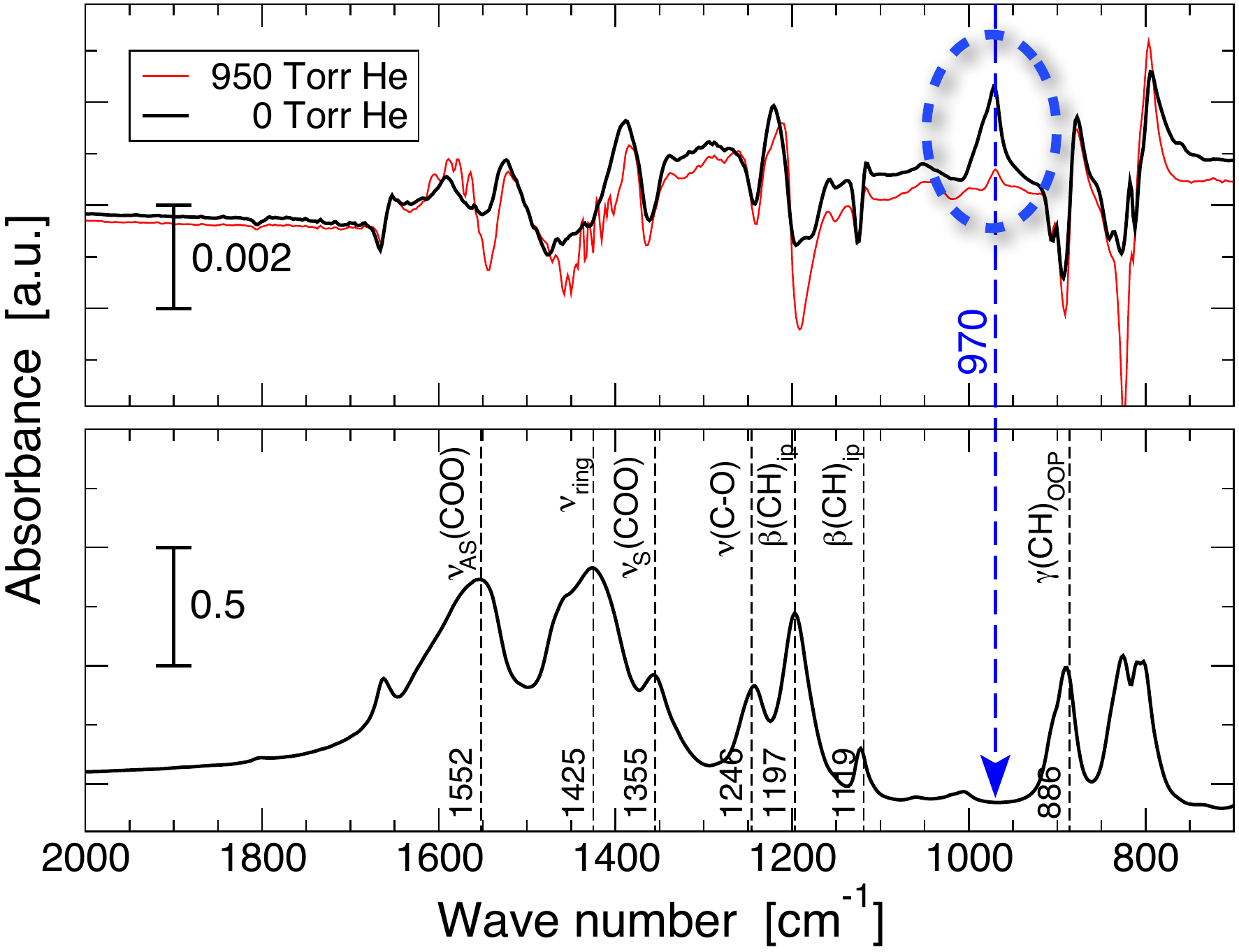}
\caption{\label{fig:IR} (bottom) IR absorption spectra of activated
Zn-{\mof} referenced to the KBr pellet. (top) IR absorption of Zn-{\mof}
after being exposed to 8~Torr of D$_2$O for 60 min.\ at 180~$^{\circ}$C,
with and without the presence of He; both spectra are referenced to the
activated sample in the bottom panel. The fingerprint of the water
dissociation reaction is clearly visible as a large peak appearing at
970~cm$^{-1}$, falling into the phonon gap of the activated sample.
Repeating the experiment with 950~Torr of He added shows that this
fingerprint disappears again.}
\end{figure}

The interaction mechanism of {\mof} with water has long been a mystery,
as no clear spectroscopic signature could be observed. The only hint
that a reaction occurred was a decrease in small-molecule uptake and the
loss of crystal structure of some {\mof} members after exposure to
water.\cite{Kizzie_2011:effect_humidity, Remy_2013:selective_dynamic,
Liu_2011:stability_effects, Schoenecker_2012:effect_water,
DeCoste_2013:effect_water} Progress in the understanding of the
interaction between water and {\mof} came only recently through infrared
(IR) absorption spectroscopy, which identified the fingerprint of a
water dissociation reaction {\reaction} at temperatures above
150~$^\circ$C.\cite{Tan_2014:water_reaction,
Tan_2015:water_interactions} This fingerprint consists of a clear peak
in the IR spectrum at 970~cm$^{-1}$, which by fortuitous coincidence
falls into the phonon gap of the MOF-74 spectrum, when exposing e.g.\
Zn-{\mof} to D$_2$O, see Fig.~\ref{fig:IR}. The corresponding peak for
the same reaction with H$_2$O is at 1316~cm$^{-1}$ and thus outside the
phonon gap, where it strongly coupling with other MOF phonon modes and
becomes impossible to detect.

The {\reaction} reaction in MOF-74 has been linked to its CO$_2$ uptake
reduction and structural instability in the presence of
water\cite{Tan_2014:water_reaction, Han_2010:molecular_dynamics} and the
exact relationship between them has just recently been
discovered.\cite{Zuluaga_2016:understanding_controlling} Therefore, if
we are able to suppress the water dissociation reaction, we are in a
position to control the reduction in gas uptake capacity and maintain
the crystal structure under humid conditions. In this work, we present
our study on how the {\reaction} reaction takes place inside {\mof} and
use our findings to design a simple but elegant experiment that shows a
means to control the reaction.  With this proof-of-principle approach,
our work lays the foundation for a concise understanding of the water
dissociation mechanism in {\mof}, opening the door for the rational
design of water stable {\mof}.  It is likely that a suppression of the
dissociation reaction can also be achieved via co-adsorption of other
molecules alongside water,\cite{Tan_2015:competitive_coadsorption} but
we focus on only water here since it binds strongest amongst many small
molecules.\cite{Canepa_2013:high-throughput_screening} For our study we
use Zn-{\mof}, as it has the highest catalytic activity amongst the
isostructural {\mof} family for water
dissociation.\cite{Tan_2014:water_reaction} Note that the catalytic
activity linearly increases as the size of the metal
decreases,\cite{Zuluaga_2016:understanding_controlling} and we find that
the catalytic activity towards the {\reaction} reaction is greatest in
Zn-{\mof}, followed by Mg-, Ni-, and Co-{\mof}, in that order.  We will
show that this behavior has its roots in the fact that---in order for
the {\reaction} reaction to take place---the water clusters need to
interact with the metal center and the linker at the same time. As the
metal center becomes smaller, it pulls the water molecules closer to the
corner of the MOF and thus enhances also the interaction of the water
with the linker.  Note that for simplicity we may use the word
``water,'' but actually refer to ``heavy water,'' i.e.\ D$_2$O.

\section{Barrier to Water dissociation}

\begin{figure}
\includegraphics[width=\columnwidth]{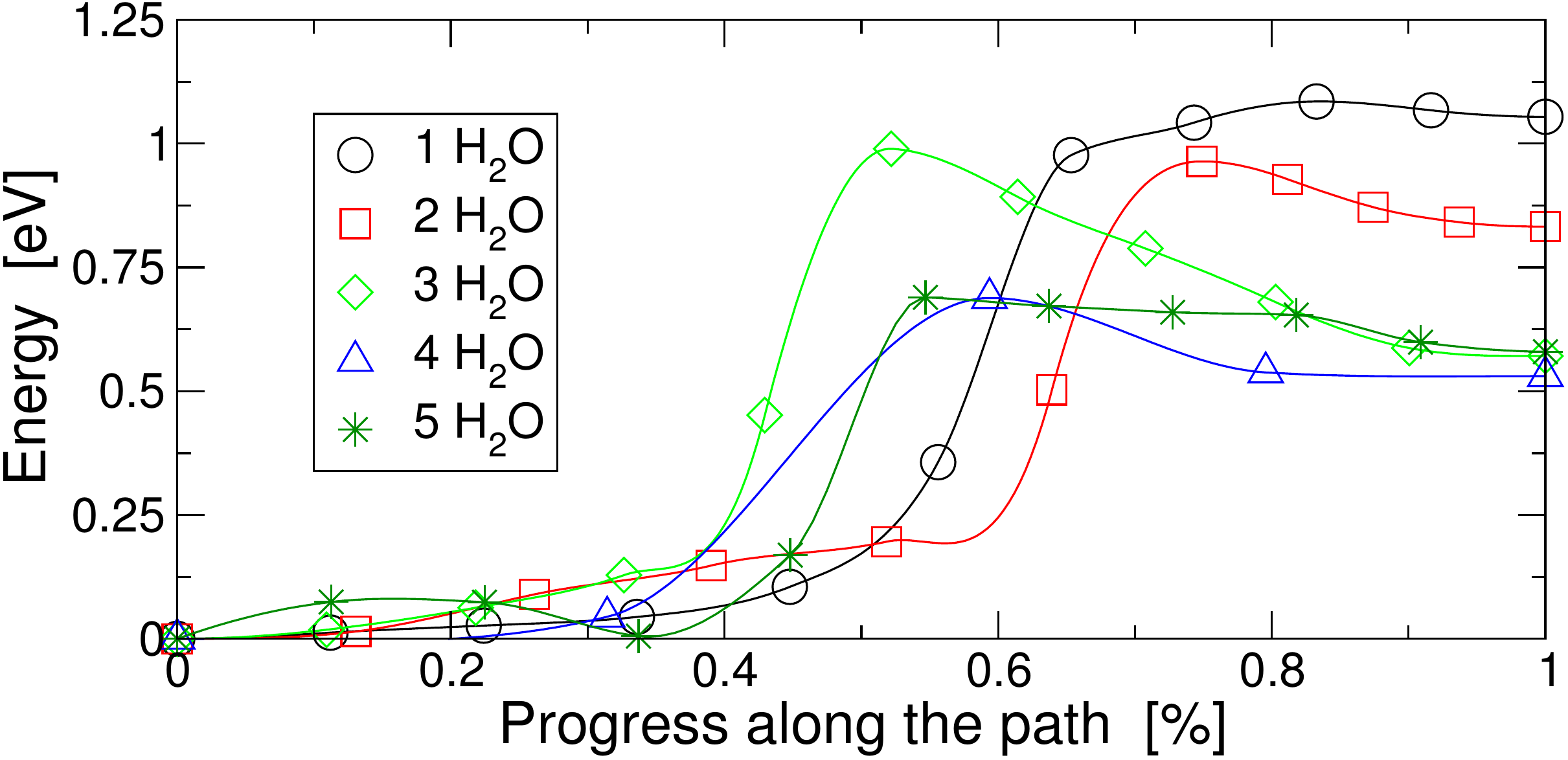}
\caption{\label{neb_fig} Relative energy [eV] along the water
dissociation reaction $n$\;H$_2$O~$\rightarrow$~OH+H+$(n-1)$\;H$_2$O  in
Zn-{\mof} for $n=1$ to $5$. The maxima for each line, i.e.\ the
activation barrier, are tabulated in Table~\ref{tab:barriers}.}
\end{figure}

\begin{figure*}
\includegraphics[width=0.4\columnwidth]{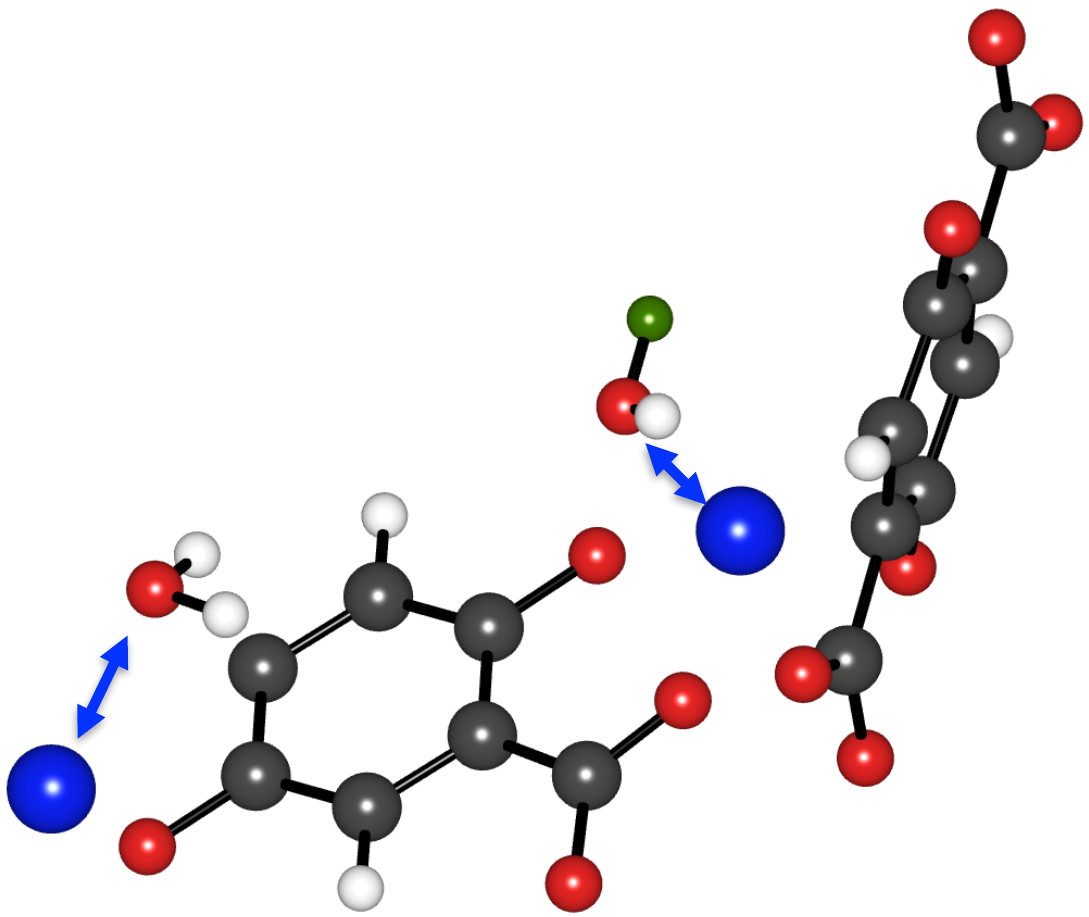}\hfill
\includegraphics[width=0.4\columnwidth]{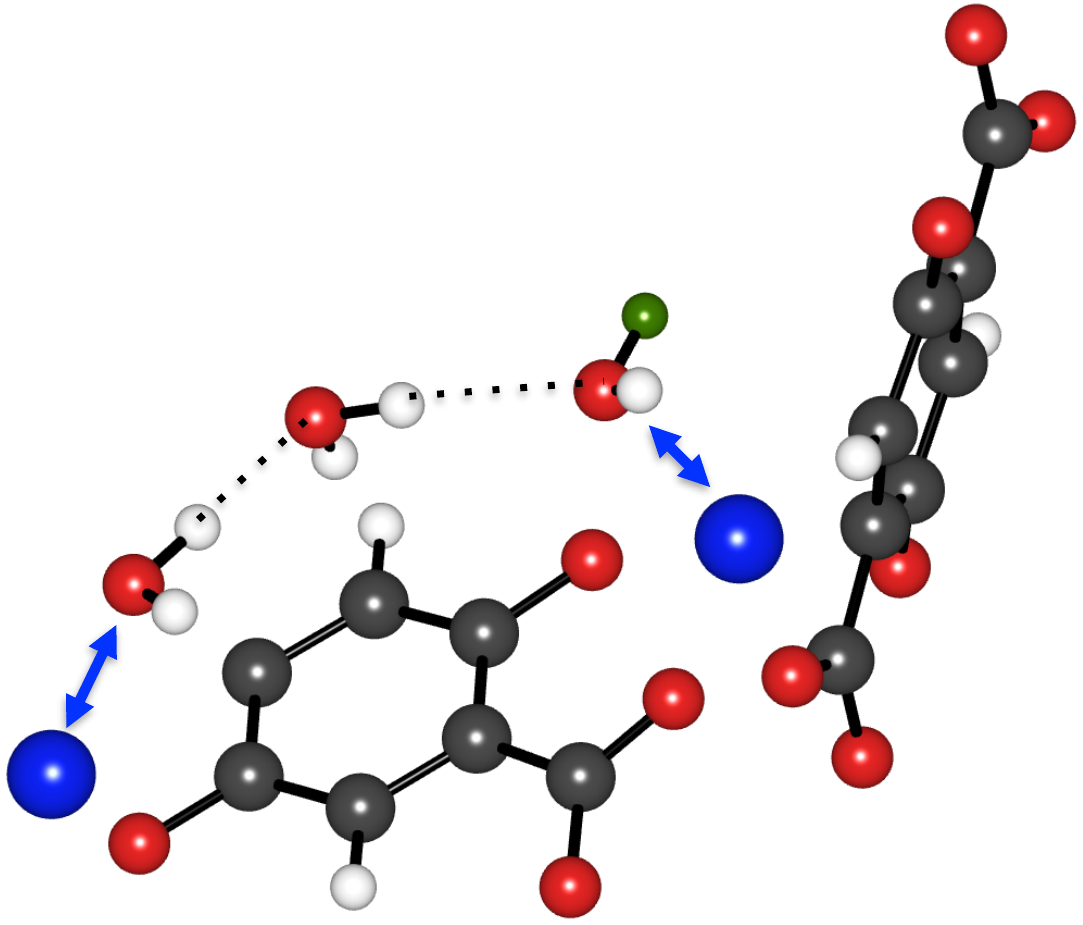}\hfill
\includegraphics[width=0.4\columnwidth]{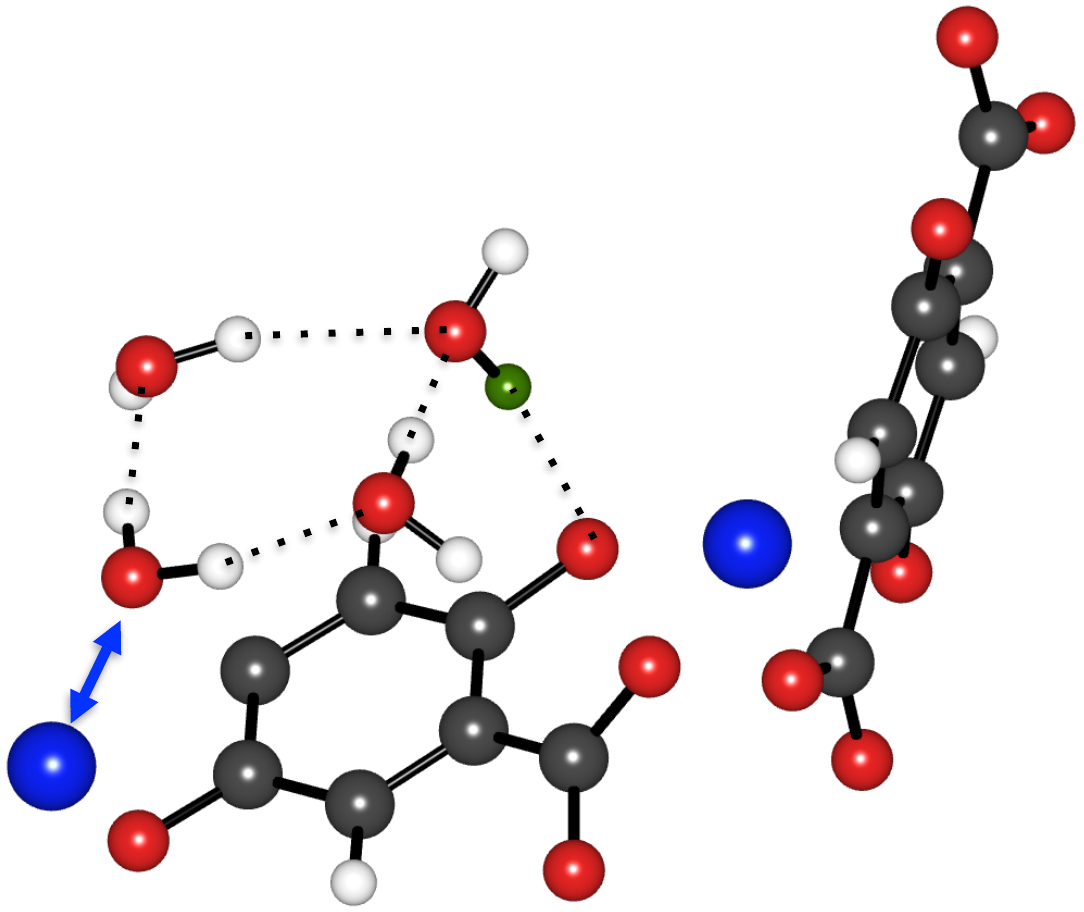}\hfill
\includegraphics[width=0.4\columnwidth]{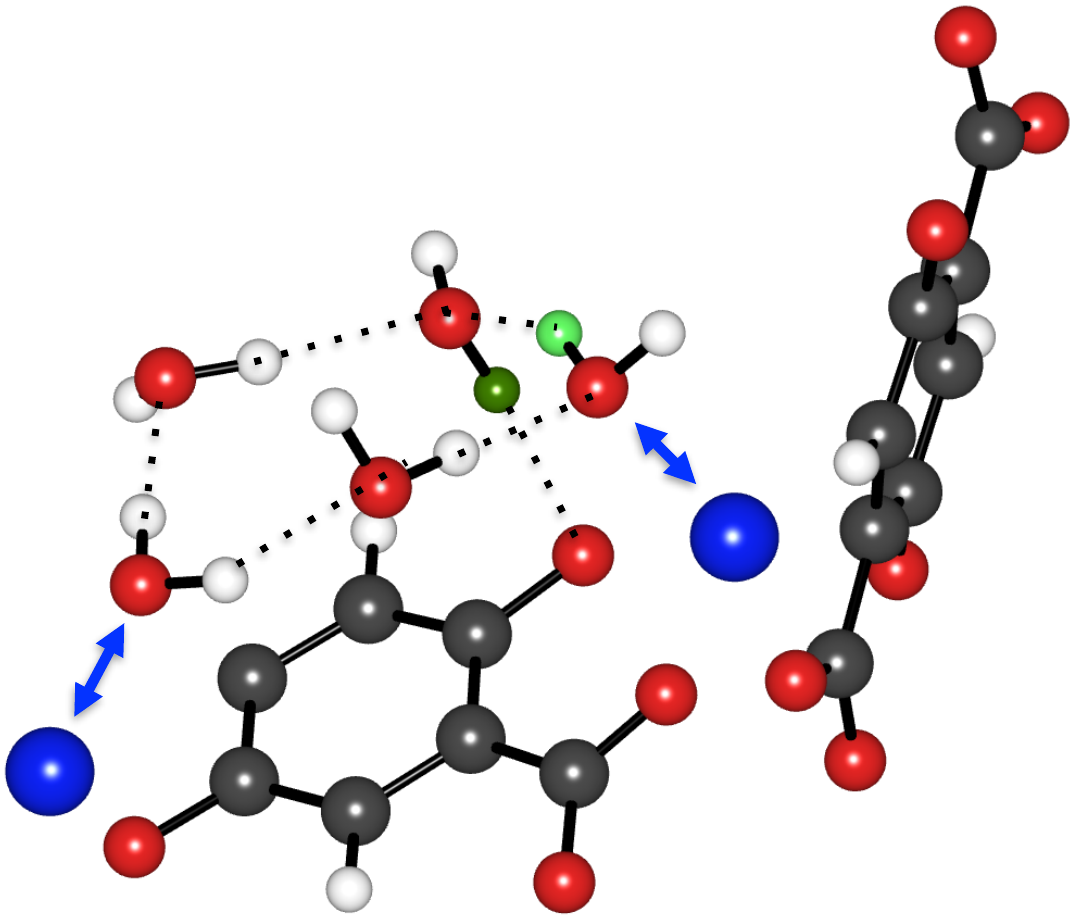}\hfill
\includegraphics[width=0.4\columnwidth]{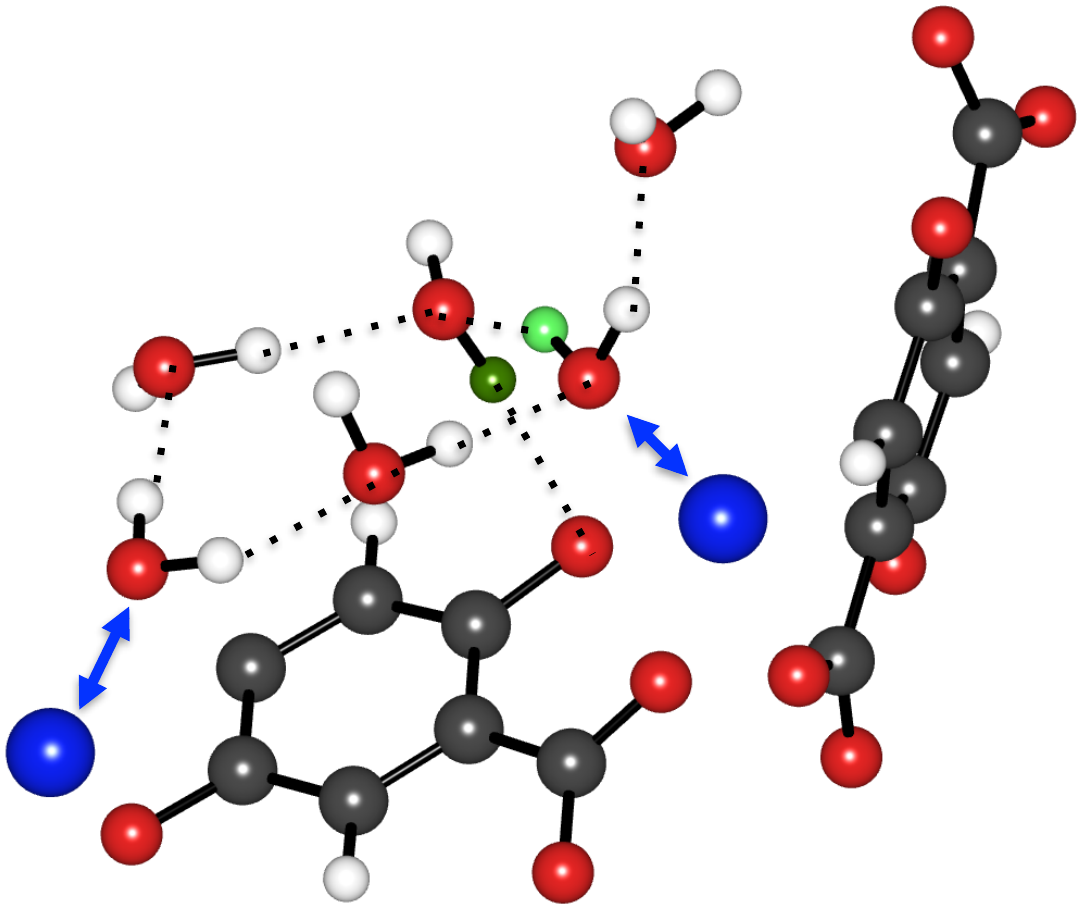}\\[2ex]
\hspace*{0.5in}$\Downarrow$\hfill\hfill
               $\Downarrow$\hfill\hfill
               $\Downarrow$\hfill\hfill
               $\Downarrow$\hfill\hfill
               $\Downarrow$\hspace{0.75in}\mbox{}\\
\includegraphics[width=0.4\columnwidth]{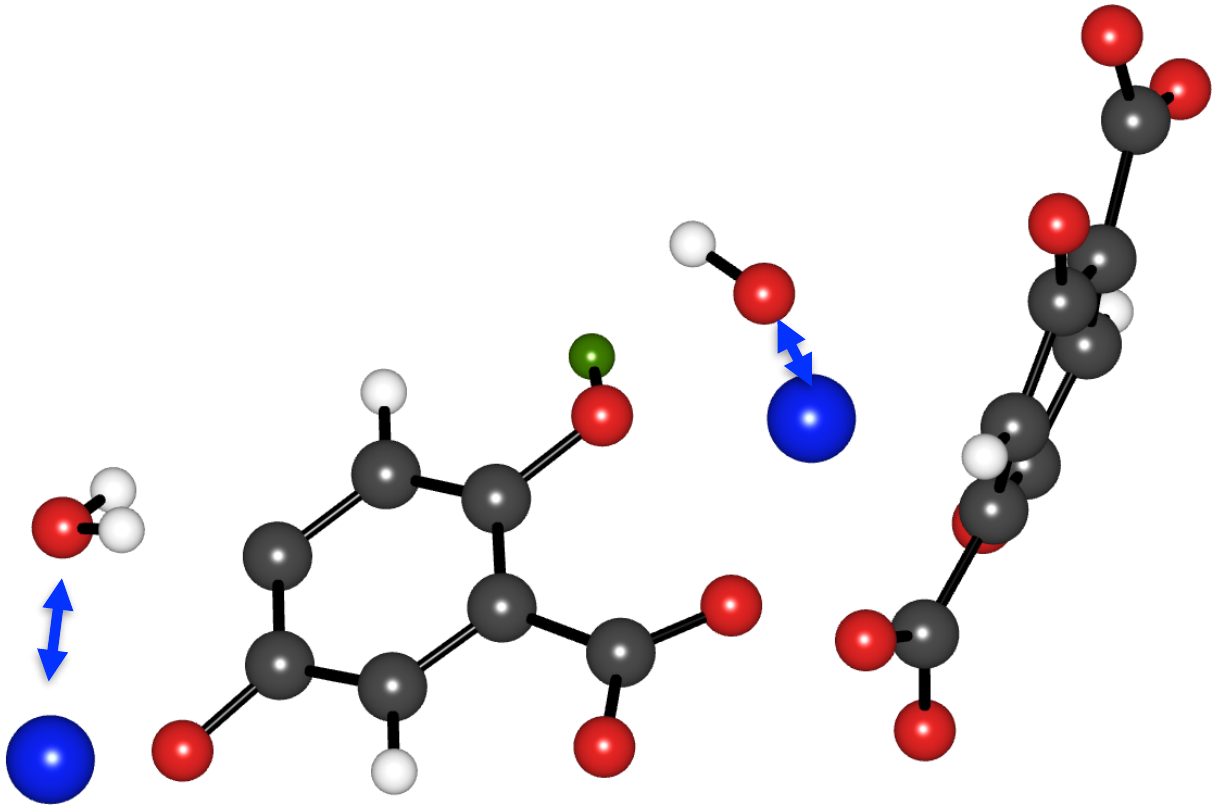}\hfill
\includegraphics[width=0.4\columnwidth]{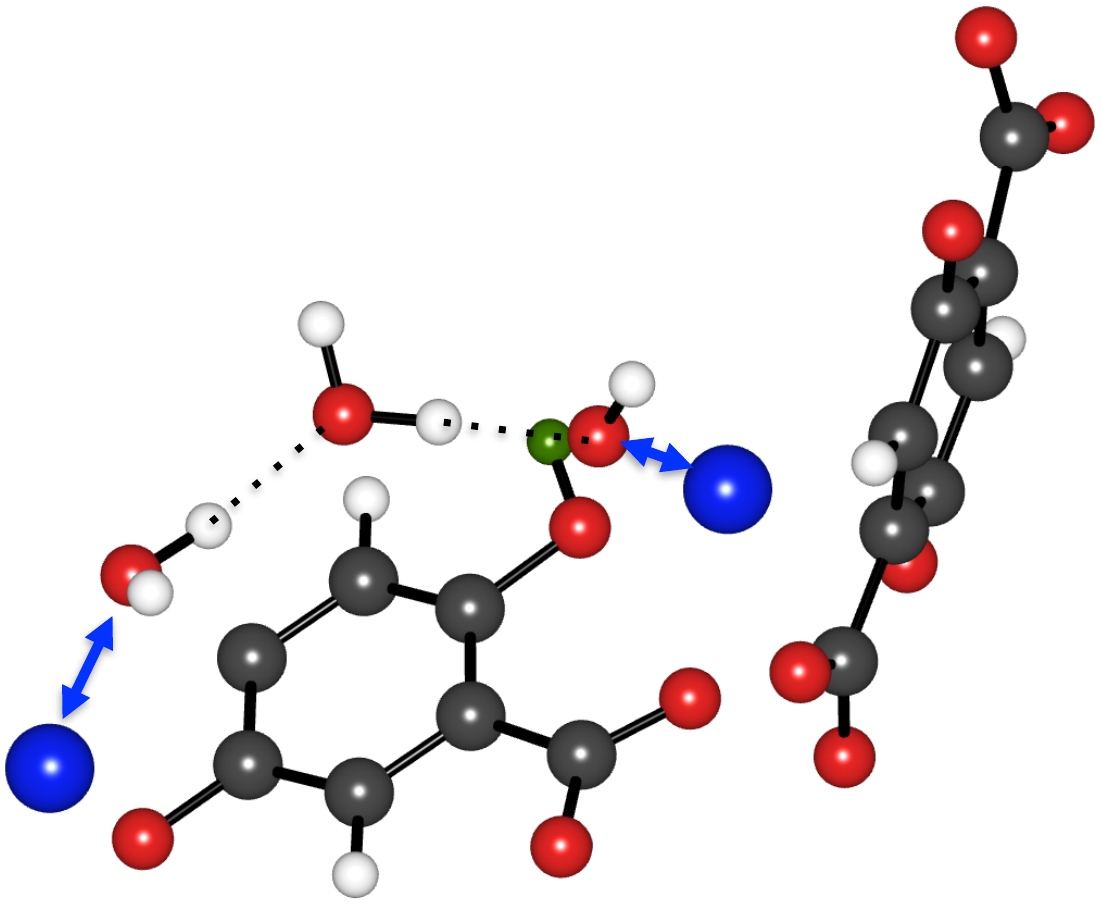}\hfill
\includegraphics[width=0.4\columnwidth]{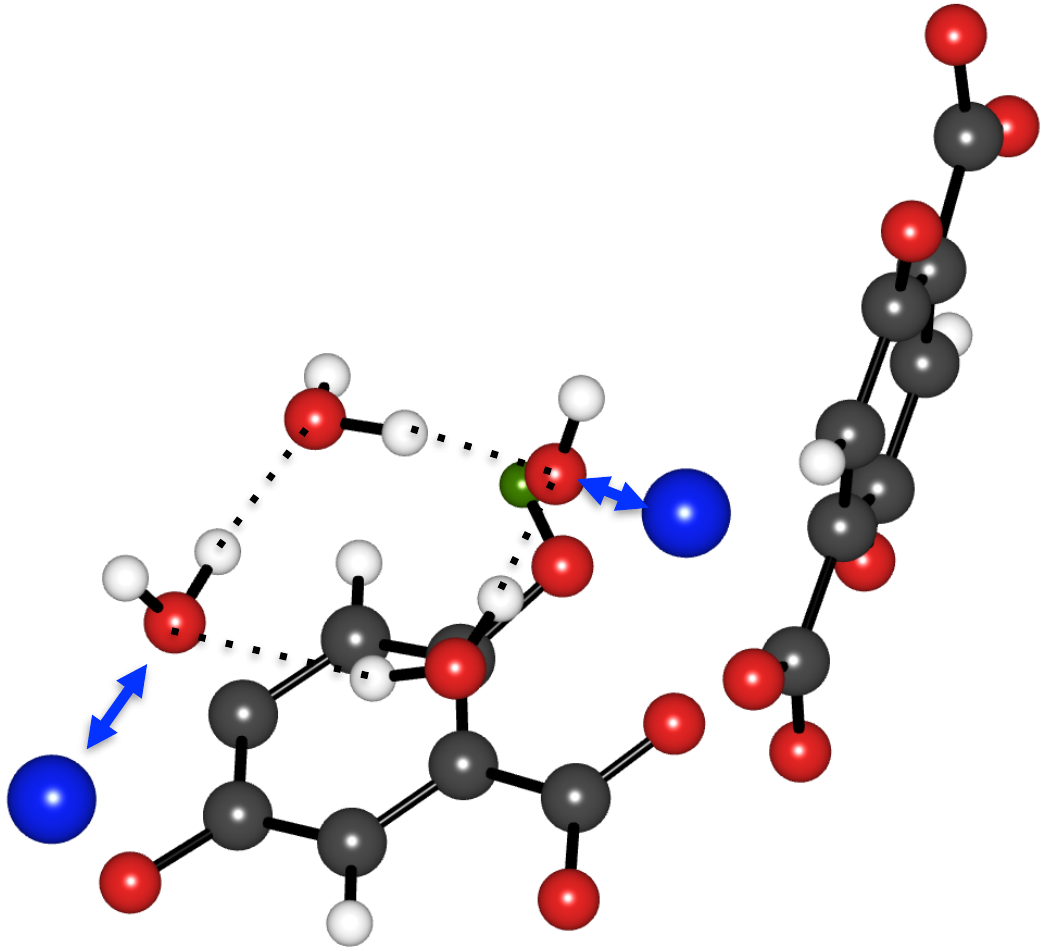}\hfill
\includegraphics[width=0.4\columnwidth]{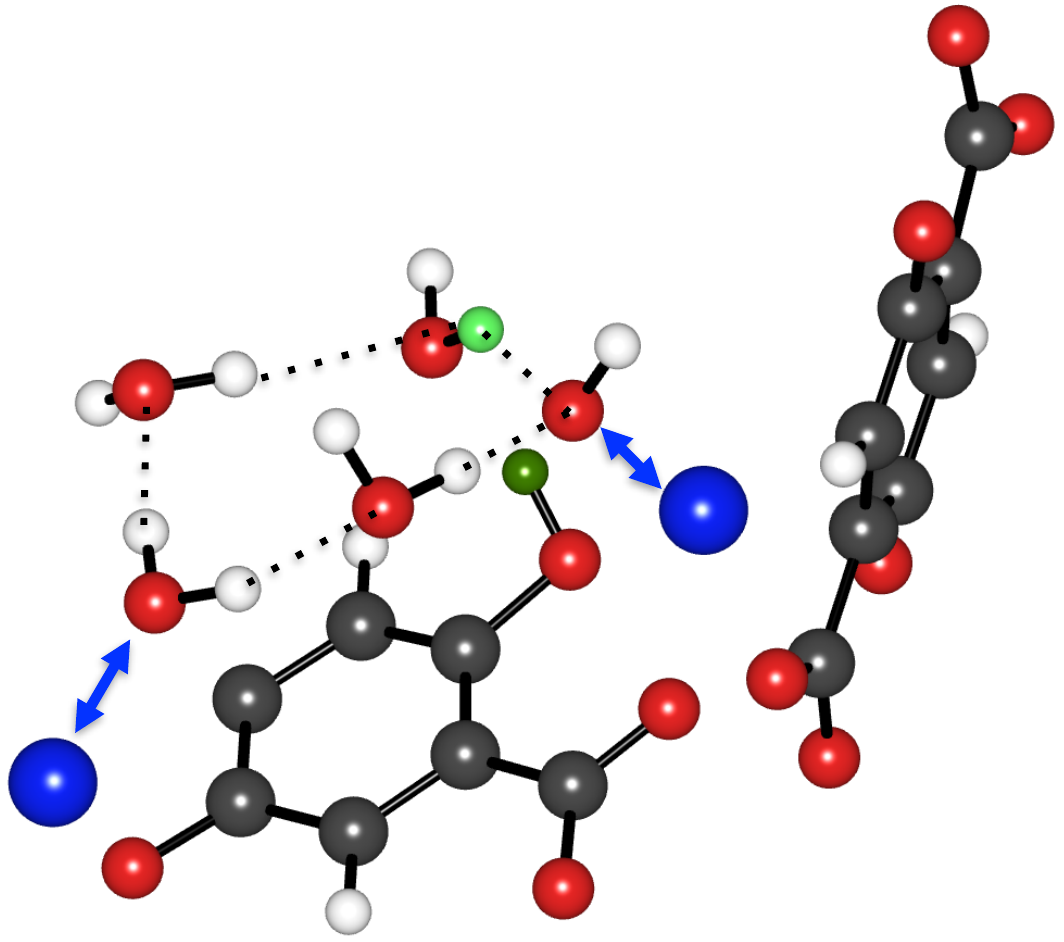}\hfill
\includegraphics[width=0.4\columnwidth]{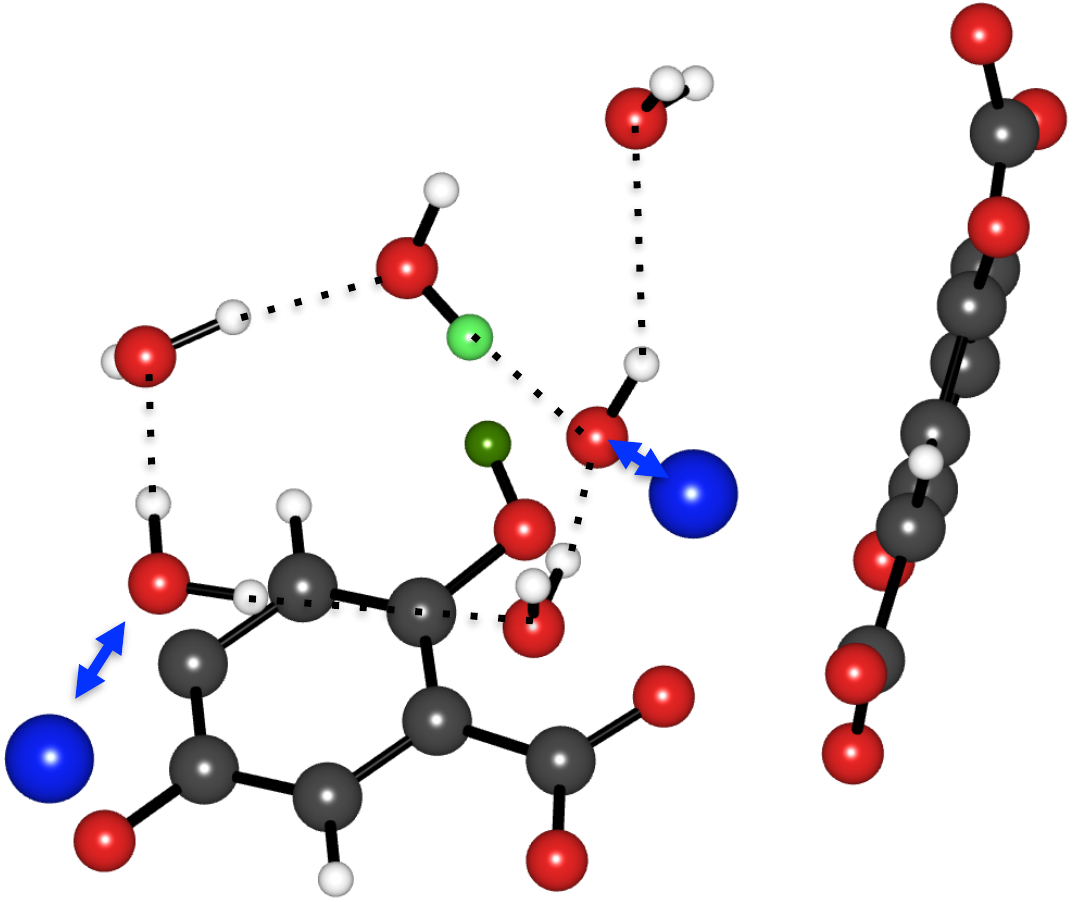}\\
\hspace*{0.5in}$n=1$\hfill\hfill
               $n=2$\hfill\hfill
               $n=3$\hfill\hfill
               $n=4$\hfill\hfill
               $n=5$\hspace{0.75in}\mbox{}
\caption{\label{fig:initial_states} Water cluster geometries for
$n=1,2,3,4$, and $5$ before (upper row) and after (lower row) the
{\reaction} reaction takes place.  Red, white, black, and blue spheres
represent O, H, C, and Zn atoms. Blue arrows indicate bonds between the
water molecules and metal centers, while dashed black lines indicate
H--O bonds. H atoms that are transferred are colored in green. The
figure only shows a small portion of {\mof}.}
\end{figure*}

We begin by using \emph{ab initio} simulations to analyze the
{\reaction} reaction at the metal centers of Zn-MOF-74 and, in
particular, how other water molecules affect it. For this purpose, we
considered $n=1$, 2, 3, 4, and 5 water molecules near one metal site,
which form small water clusters.  The energy profile along the
{\reaction} reaction pathway and its thermodynamics for the various
clusters are given in Fig.~\ref{neb_fig}; actual values for activation
barriers ($E_B$) are collected in Table~\ref{tab:barriers}. For a single
water molecule cluster ($n=1$), we find a reaction barrier of 1.09~eV.
The {\reaction} reaction path corresponds to
Fig.~\ref{fig:initial_states}, where the adsorbed water transfers one
hydrogen to the nearest oxygen of the linker, as discussed in detail
elsewhere.\cite{Tan_2014:water_reaction} The stable states of these
clusters before the dissociation reaction are depicted in the upper row
of Fig.~\ref{fig:initial_states}; the lower row shows their counter
parts, once the dissociation reaction has taken place. Note that for
large enough clusters, the clusters start interacting with one
additional water molecule at the nearby metal site and we have not
included that molecule in our count of $n$, i.e.\ there are $n+1$ water
molecules in each figure. The water molecules in the clusters interact
via H bonds, similar to their behavior on the surface of metals and
other systems.\cite{Carrasco_2012:molecular_perspective,
Maheshwary_2001:structure_stability, Nie_2010:pentagons_heptagons,
Feibelman_2010:interpretation_high-resolution,
Henderson_2002:interaction_water, Kolb_2011:van_waals,
Meyer_2004:partial_dissociation}

\begin{figure*}[t]
\hspace*{\fill}\includegraphics[width=0.5\columnwidth]{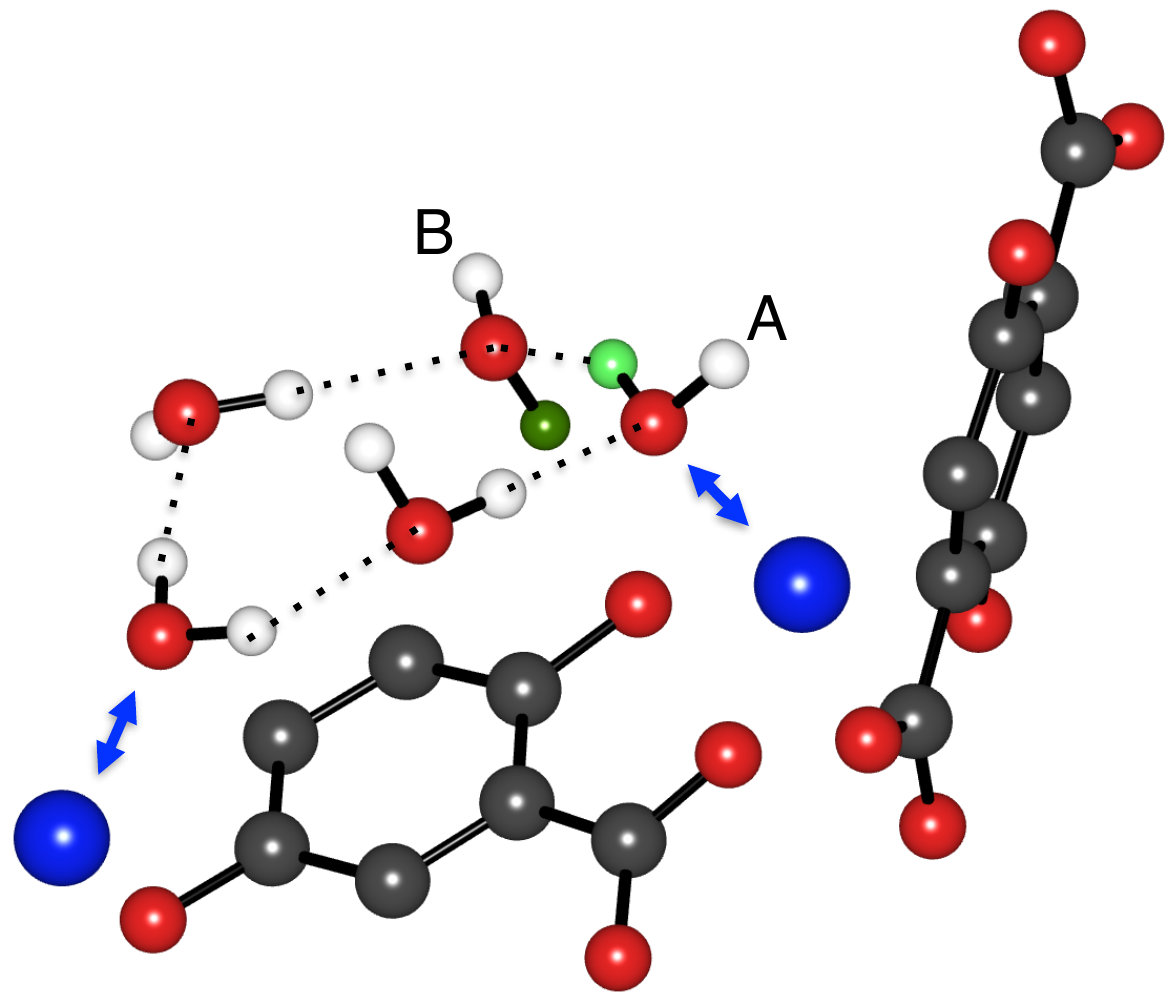}\hfill
\includegraphics[width=0.5\columnwidth]{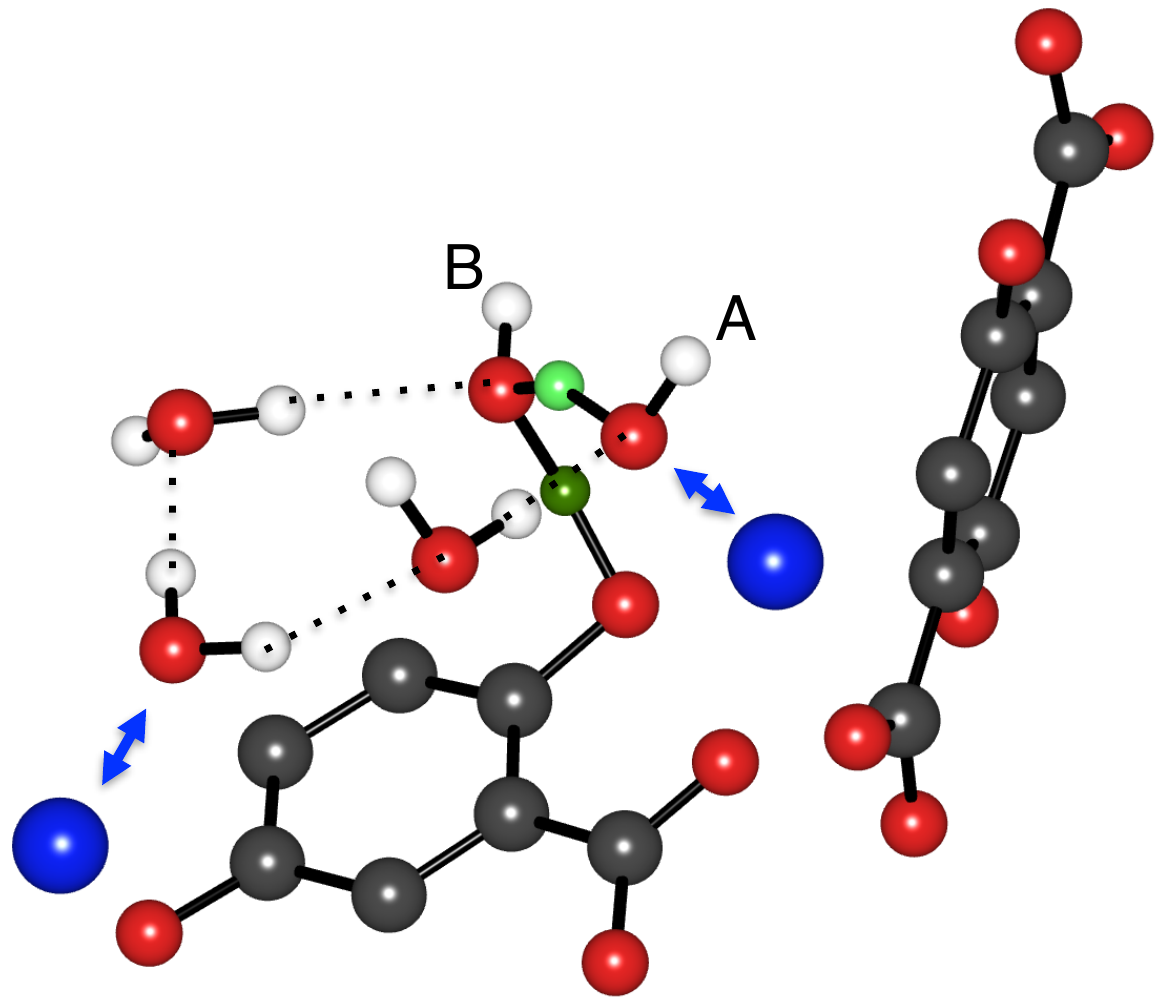}\hfill
\includegraphics[width=0.5\columnwidth]{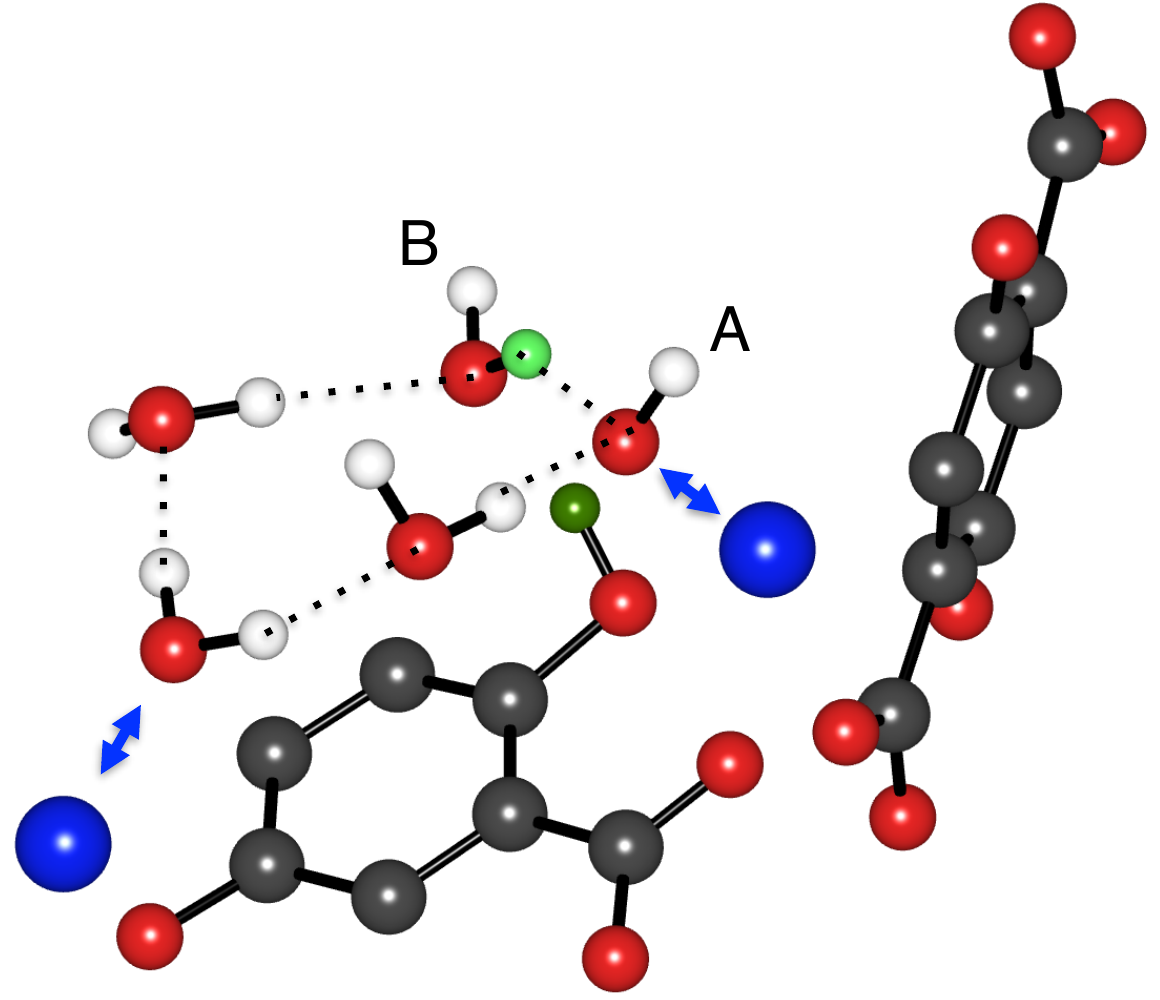}\hspace*{\fill}\mbox{}
\caption{\label{fig:reaction_pathway}Initial, transition, and final
state of the water dissociation pathway corresponding to $n=4$. The
reaction takes place through the following low-energy barrier path:
\atomA\ donates one H to \atomB, which---at the same time---donates one
H to the oxygen of the linker. Red, white, black, and blue spheres
represent O, H, C, and Zn atoms; the hydrogens being transferred are
shown in green. See the corresponding animation in the supplementary
materials.  The figure only shows a small portion of {\mof}.}
\end{figure*}

Looking at the barriers for water clusters with $n=1$ to 3, we see that
they only decrease slightly, see Table~\ref{tab:barriers} and
Fig.~\ref{neb_fig}. In these cases, the additional water molecules act
as catalysts and the reaction proceeds just like in the case of a single
water molecule---i.e.\ the water molecule at the metal center donates
its H to the oxygen of the linker, see Fig.~\ref{fig:initial_states}.
Other groups have suggested a similar effect on metal oxide surfaces,
called water catalyzed dissociation induced by collective
interaction,\cite{Tocci_2014:olvent-induced_proton,
Hass_1998:chemistry_water, Giordano_1998:partial_dissociation,
Odelius_1999:mixed_molecular} where the interactions among water
molecules play an important role, responsible for a change in the
structure and proton-transfer dynamics. However, in our case the
interaction among water molecules plays an additional role, as we show
next: for  $n>3$ we observe a dramatic 37\% reduction in the activation
barrier, see Table~\ref{tab:barriers} and Fig.~\ref{neb_fig}. This
surprising and unexpected finding is the result of the water
dissociation now proceeding through a completely different pathway. For
$n$=4 and 5 the geometry of the cluster allows for one of the water
molecules to position above the linker, establishing a H bond with the O
atom of the linker. This geometry opens up a much more favorable
pathway. For example, Fig.~\ref{fig:reaction_pathway} shows the
{\reaction} reaction path for the $n=4$ case: Water molecule \atomA\
donates one H to molecule \atomB, which, at the same time donates one H
to the oxygen at the linker---an animation of this reaction is available
in the supplementary materials.  We have also investigated the case of
$n=5$, in which the reaction proceeds similarly to the $n=4$ case.  Note
that this is vastly different from the cases of $n=1$ to 3, where the H
being donated to the linker comes from the water adsorbed at the nearby
metal center.

\begin{table}
\caption{\label{tab:barriers} Activation barrier $E_B$ [eV] for the
water dissociation reaction
$n$\;H$_2$O~$\rightarrow$~OH+H+$(n-1)$\;H$_2$O at one metal center in
Zn-{\mof}. Figure~\ref{neb_fig} shows the energy along the entire
reaction pathway. We also list formation energies per water molecule
$E_F$ [eV] for all clusters.}
\begin{tabular*}{\columnwidth}{@{}l@{\extracolsep{\fill}}rrrrr@{}}\hline\hline
$n$   &      1  &       2 &       3 &       4 &      5 \\\hline
$E_B$ &   1.09  &    0.96 &    0.99 &    0.69 &   0.69 \\
$E_F$ & $-0.57$ & $-0.57$ & $-0.61$ & $-0.62$ & $-0.62$\\\hline\hline
\end{tabular*}
\end{table}

\begin{figure}
\centering\includegraphics[width=0.6\columnwidth]{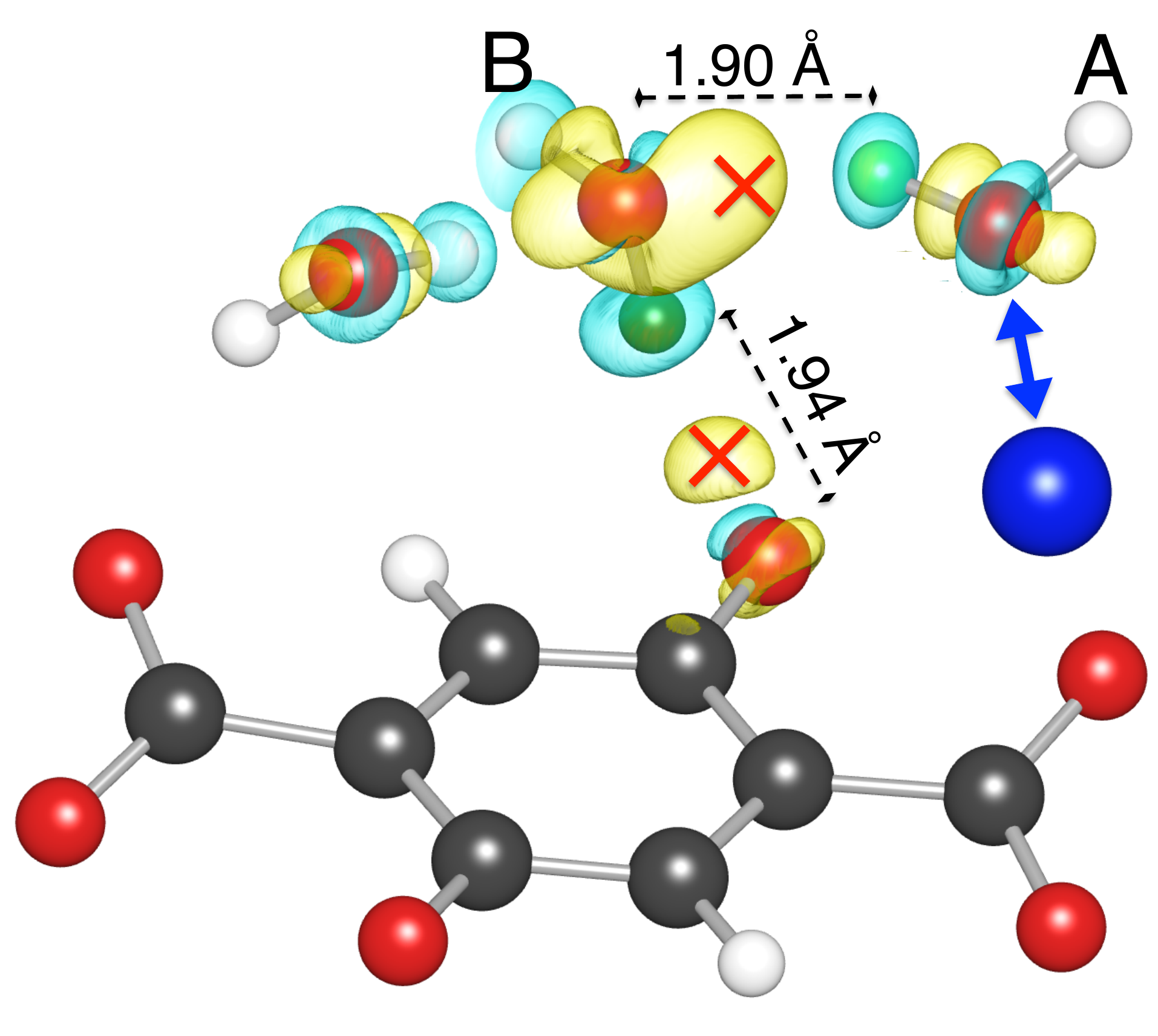}
\caption{\label{fig:charge}Charge density redistribution upon adsorption
of water molecule \atomB\ for $n=4$. Yellow denotes charge accumulation,
while blue denotes charge depletion; the iso level is
$\pm$0.003~$e$/\AA$^3$. The red $\times$ marks strong bonds between
\atomA\ and \atomB\ and between \atomB\ and the linker. For clarity,
some water molecules and other elements of the system are not shown in
this figure. Color coding as in Fig.~\ref{fig:initial_states}.}
\end{figure}

The key to this drastic lowering of the energy barrier for $n>3$ is the
water molecule above the linker. The {\reaction} reaction can proceed
through this low energy-barrier path only if one of the water molecules
in the cluster has a strong interaction with the O of the linker. In
other words, one water molecule needs to be close to the O of the
linker. The top row of Fig.~\ref{fig:initial_states} shows the
configuration of the water molecules before the {\reaction} reaction
takes place for $n=1,2,4$, and $5$. For $n=4$ and $5$, we find that one
of the water molecules is able to position itself above the linker and
establish an H bond with the O of the linker, while for $n=1$ and $2$
this is not possible. For comparison, in the $n=1$ case the water
molecule adsorbed on the metal center is located 3.64~\AA ~away from the
O of the linker, whereas for the $n=4$ case the water molecule is
located only 1.94~\AA\ away. In Fig.~\ref{fig:charge} we analyze the
role of this special water molecule near the O of the linker further.
This figure shows the charge density redistribution upon adsorption of
the water molecule \atomB. We find significant charge accumulation
(marked as red $\times$ marks) between molecules \atomA\ and \atomB, and
between molecule \atomB\ and the linker. These strong bonds are
responsible for the short distance between \atomA\ and \atomB\ (1.90
\AA) and between \atomB\ and the O of the linker (1.94 \AA), allowing
for the significantly reduced reaction barrier. Now also the function of
the other water molecules in the clusters for $n=4$ and $5$ becomes
apparent---the remaining water molecules forming the cluters serve to
restrain \atomB\ in its particular geometry above the O of the linker,
making the close interaction with the linker and the reduced energy
barrier possible.

Note that, even though for $n=3$ one water molecule is close to the O of
the linker (see Fig.~\ref{fig:initial_states}), the energy barrier for
the reaction is still 0.99~eV (see Table~\ref{tab:barriers}). The reason
why the barrier is not reduced in this particular case is that, once the
H$_2$O molecule loses its H, the remaining OH has to travel a long
distance to the metal center in order to find a stable state.

Before we move on to the experimental verification of our discovered
water dissociation mechanism, we would like to comment on two aspects.
First, water dissociation in \mof\ takes place at temperatures above
150~$^\circ$C.\cite{Tan_2014:water_reaction} Water clusters of such size
as we are considering here may not be stable at such temperatures in
their gas phase. However, the clusters inside \mof\ are stabilized
through interaction with the linker. To quantify this aspect, we have
calculated the formation energies per water molecule $E_F$ for all
clusters, reported in Table~\ref{tab:barriers}. The numbers for all
clusters inside \mof\ ($\sim-0.6$ eV) indicate a significantly higher
stability than the corresponding number of $-0.32$~eV for their
gas-phase equivalents.\cite{Kolb_2011:van_waals} Second, due to the
nature of transition-state search algorithms, the energy barriers
reported in this work are an upper limit.  There is a large number of
stable initial and final configurations for water clusters that give
rise to numerous possible reaction pathways.  While we have modeled
approximately 50 reactions, we only report the paths with the lowest
energy barriers. For $n>5$ there may be other geometries that exhibit
yet a lower energy barrier for the {\reaction} reaction. These paths may
even have energy barriers closer to the splitting of water in other
systems, where the energy barriers are as low as
$\sim$0.1~eV.\cite{Lei_2007:initial_interactions,
Kostov_2005:dissociation_water, Tilocca_2003:reaction_pathway,
Miao_2012:activation_water} However, a comprehensive \emph{ab initio}
sampling of larger clusters is prohibitively expensive. But, our main
goal is not to find such paths or ways to dissociate water in {\mof} in
the most efficient manner. On the contrary, our goal is to understand
the underlying mechanism and use this knowledge to avoid the {\reaction}
reaction inside {\mof}. If we are able to do so, we would be able to
avoid the poisoning of the metal centers (going hand-in-hand with a
decrease of the small-molecule uptake
capacity)\cite{Tan_2014:water_reaction} and the degradation of the
crystal structure of the {\mof} in the presence of
water.\cite{Han_2010:molecular_dynamics,
Zuluaga_2016:understanding_controlling} Our results suggest a way to
accomplish exactly this goal: If we avoid the clustering of water
molecules around the metal centers, we increase the energy barrier for
the {\reaction} reaction from 0.69~eV (or lower) up to 1.09~eV, and the
reaction will not take place at the temperatures at which {\mof} is
stable, i.e.\ between 0 and 300~${^\circ}$C. In the next section we show
a way of achieving this.

\section{Controlling water dissociation using He}

Based on our theoretical results, we hypothesize that, if we are able to
prevent the formation of these water clusters on the walls of {\mof}, we
will be able to control and even suppress the water dissociation
reaction, with important effects on the MOF's stability and
small-molecule uptake capacity. To test our hypothesis, we have designed
an experiment where we control the formation of water clusters by
introducing a non-reactive agent into the {\mof} channel, such as He.
More specifically, in addition to 8~Torr of D$_2$O we now introduce also
500 and 950~Torr of He into the activated Zn-{\mof} at 180~$^{\circ}$C,
see the experimental details section for further details. The role of
these non-interacting He atoms is to get in-between the water molecules,
decreasing their interaction, preventing hydrogen bonds, and thus
hindering the formation of clusters.  At the same time, the partial
pressure of He provides an effective means of controlling the number of
He atoms in each nano pore and thus the degree of water cluster
formation. This directly affects the reaction barrier for the
dissociation reaction and allows us to control the amount of water
dissociated.

\begin{figure}[t]
\includegraphics[width=\columnwidth]{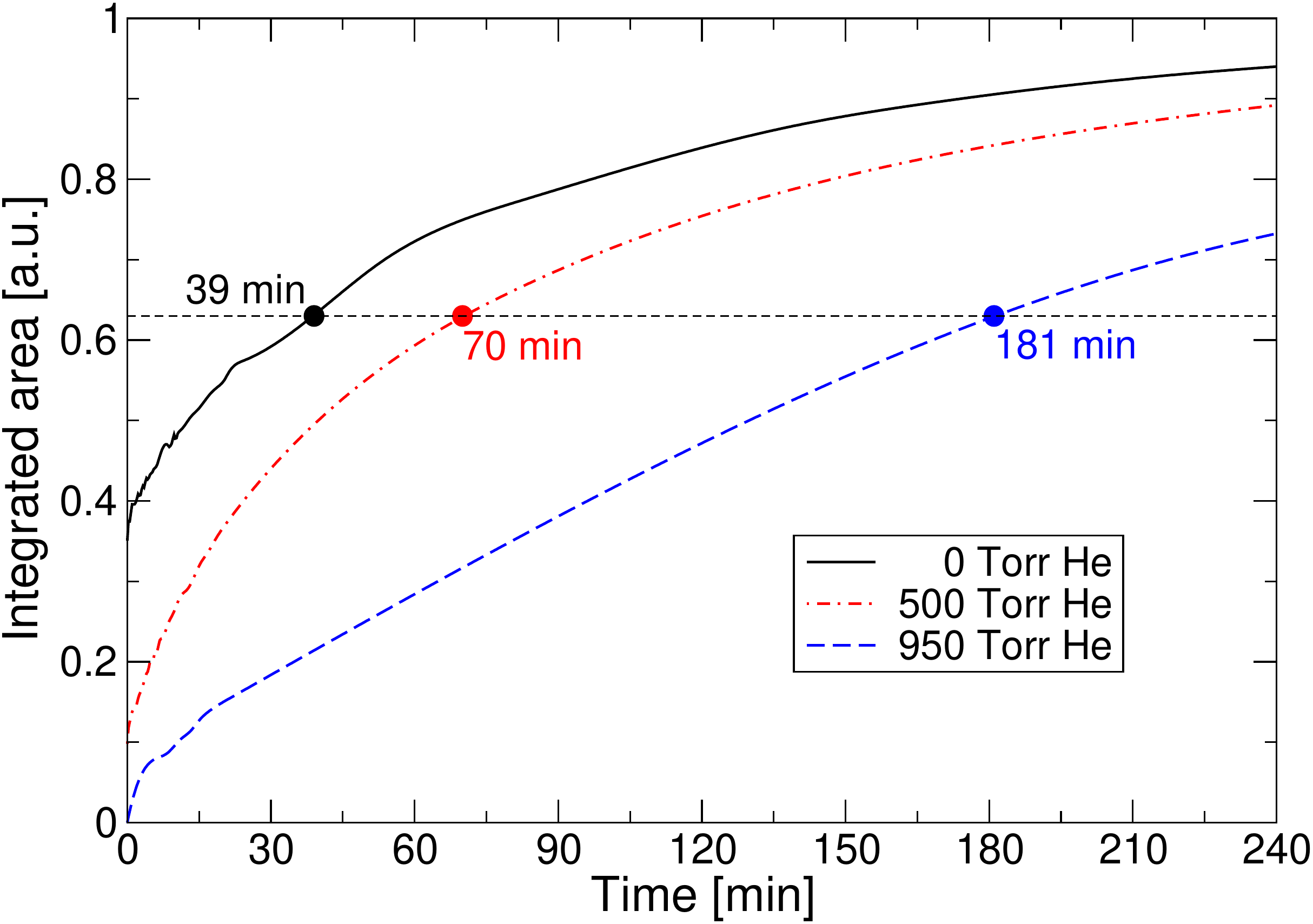}
\caption{\label{fig:integrated_area}Integrated area of the normalized
peak at 970~cm$^{-1}$ from Fig.~\ref{fig:IR} as a function of time,
giving a quantitative measure of the amount of water dissociated. All
samples are exposed to 8~Torr of D$_2$O and a temperature of
180~$^{\circ}$C.  The black curve corresponds to the sample with no
additional He, while the red, and blue curves correspond to the samples
exposed to 500 and 950~Torr of He, respectively. The plot confirms that
the water dissociation reaction is suppressed as higher partial
pressures of He are introduced.} 
\end{figure}

For reasons explained in the introduction, our experiments are performed
with D$_2$O instead of H$_2$O. The {\reactionD} reaction has a
characteristic fingerprint in form of a clear peak in the IR spectrum at
970~cm$^{-1}$ (see Fig.~\ref{fig:IR}), the integrated area of which
provides a quantitative measure of how much water has dissociated. Our
experiment now proceeds to record the integrated area of this peak
(normalized to~1)  as a function of time for the three different He
partial pressures; results are plotted in
Fig.~\ref{fig:integrated_area}.  As can be seen, the integrated area for
the sample without He (0~Torr He) grows and saturates as the water
molecules on the metal centers dissociate; the reaction reaches 63\% of
the saturation level in 39~minutes. However, as the He partial pressure
increases, the water dissociation reaction is suppressed due to the
higher activation barrier, as the water clusters are less likely to
form.  In turn, we find that for He partial pressures of 500 and
950~Torr, it takes 70 and 181 minutes to reach 63\% of the saturation
level, respectively, confirming our hypothesis and the cluster assisted
water dissociation mechanism. As an example, the top panel of
Fig.~\ref{fig:IR} shows the decrease of the peak at 970~cm$^{-1}$ after
60 minutes when adding 950~Torr of He to the 8 Torr of D$_2$O.

We have also used non-reactive agents other than He to demonstrate the
connection between cluster formation, reaction barrier, and water
dissociation suppression. Our results indicate that, e.g.\ the
introduction of Ar also reduces the rate at which the water dissociation
reaction takes place. However, the effect is not as strong as for He.
Using 950~Torr of Ar together with 8~Torr of D$_2$O at a temperature of
180~$^{\circ}$C, we find that the {\reaction} reaction reaches 63\% of
the saturation level after 92 minutes.  There is thus no particular
requirement for the agent, besides being non-reactive, to suppress the
{\reaction} reaction. Nevertheless, He results were chosen here as the
phenomena is stronger compare to Ar and others.  We believe that He is a
better reaction moderator because of its small size; it is fairly easy
for He to penetrate in-between water molecules in the water clusters,
suppressing their interaction and formation.

The partial pressures of He used in our experiments are large compared
to the partial pressure of water. As such, it is not practical to use He
to suppress the formation of water clusters. However, we see our
experimental results not necessarily as a practical way in large
applications to control the water dissociation, but rather as a
proof-of-principle that such control is achievable. Furthermore, our
experiments function also as crucial support for our \emph{ab initio}
results about how small water cluster formation aids the water
dissociation. In future research, the design of a water stable MOF-74
has to begin by finding a way of suppressing the water cluster formation
and the water dissociation reaction.

\section{Summary}

Our \emph{ab initio} results show that the formation of water clusters
on the walls of Zn-{\mof} allows for an energetically much more
favorable water dissociation path for the {\reaction} reaction. We then
confirm via \emph{in situ} IR spectroscopy that introducing He into the
MOF channel to prevent the formation of such clusters is an effective
way to suppress the water dissociation reaction. While we show results
explicitly for He, it is quite likely that other gases have the same
effect. Since previous studies have indicated that the water
dissociation reaction reduces MOF-74 gas uptake and crystal stability,
we have thus found a proof-of-principle approach to control them. With
this new insight, our results provide new opportunities for designing
water stable {\mof}, focusing on the suppression of the water
dissociation reaction which causes the problems.

\section{Methods}

\subsection{Computational Details}

Our \emph{ab initio} results were obtained at the density functional
theory level, as implemented in \textsc{quantum
espresso}.\cite{Giannozzi_2009:quantum_espresso} In order to correctly
capture the crucial van der Waals interactions between the MOF and the
water molecules, as well as between water
molecules,\cite{Kolb_2011:van_waals} we used the truly non-local
functional vdW-DF.\cite{Thonhauser_2015:spin_signature,
Berland_2015:van_waals, Langreth_2009:density_functional,
Thonhauser_2007:van_waals} Nuclear quantum effects, which can play a
noticeable role in the description of water and water
dissociation,\cite{McMahon_2014:combined_influence} were not taken into
account due to the extreme computational cost.  Ultrasoft pseudo
potentials were used with cutoffs of 544~eV and 5440~eV for the wave
functions and charge density, respectively. Due to the large dimensions
of the unit cell, only the $\Gamma$ point was sampled. Reaction barriers
were found with a transition-state search algorithm, i.e.\ the
climbing-image nudged-elastic band
method,\cite{Henkelman_2000:climbing_image,
Henkelman_2000:improved_tangent} using up to 11 images depending on the
path. We started from the experimental rhombohedral structure of
Zn-{\mof} with 54 atoms in its primitive cell and space group
R$\bar{3}$. The rhombohedral axes are $a=b=c=15.105$~\AA\ and
$\alpha=\beta=\gamma=117.78^\circ$.\cite{Zhou_2008:enhanced_h2} We
optimized all atomic positions until the forces were less than
$2.6\times10^{-4}$~eV/\AA.

\subsection{Experimental Details}

Zn-MOF-74 powder ($\sim$2mg) was pressed onto a KBr pellet ($\sim$1cm
diameter, 1--2~mm thick). The sample was placed into a high-pressure
high-temperature cell (product number P/N 5850c, Specac Ltd, UK) at the
focal point compartment of an infrared spectrometer (Nicolet 6700,
Thermo Scientific, US). The samples were activated under vacuum ($<$
20~mTorr) at 180~${^\circ}$C for at least 4~hour. A mixture of He and
D$_2$O was prepared in a compartment connected but separated from the
main cell.  Different ratios between He and D$_2$O were introducer into
the main cell held at 180~${^\circ}$C: 950~Torr He/8~Torr D$_2$O;
500~Torr He/8~Torr D$_2$O;  and 0 Torr He/8~Torr D$_2$O. The area under
the peak at 970~cm$^{-1}$, which is a quantitative measure of the amount
of water dissociated in Zn-{\mof}, was measured as a function of time.
Spectra were recorded until the reaction was saturated.

\section*{Acknowledgements}

This work was supported by the Department of Energy Grant No.\
DE--FG02--08ER46491. It further used resources of the Oak Ridge
Leadership Computing Facility at Oak Ridge National Laboratory, which is
supported by the Office of Science of the Department of Energy under
Contract DE--AC05--00OR22725.

\bibliography{references,biblio}

\end{document}